\documentclass[10pt,aps,prd,twocolumn,notitlepage,nofootinbib,floatfix,superscriptaddress]{revtex4-1}

\usepackage{amsmath, amsthm, amssymb, amsfonts, amsbsy, mathrsfs}

\usepackage{xcolor}
\usepackage{calligra,bm}
\usepackage{stmaryrd}

\usepackage{graphicx}
\graphicspath{{./images/}}

\usepackage[hidelinks,bookmarks=true]{hyperref} 
\hypersetup{pdfstartview=FitH,pdfhighlight=/O,colorlinks=false}

\begin{document}

\title{Multiscale feature integration network for inpainting of full-sky CMB $B$-modes}
\author{Reyhan D. Lambaga}
\email{d09244004@ntu.edu.tw}
\affiliation{Graduate Institute of Astrophysics, National Taiwan University, Taipei 10617, Taiwan}
\affiliation{Department of Physics and Center for Theoretical Sciences, National Taiwan University, Taipei 10617, Taiwan}
\affiliation{Leung Center for Cosmology and Particle Astrophysics, National Taiwan University, Taipei 10617, Taiwan}
\author{Vipin Sudevan}
\email{vipinsudevan1988@gmail.com}
\affiliation{Department of Physics, National Sun Yat-sen University, No. 70, Lien-Hai Road, Kaohsiung City 80424, Taiwan, R.O.C.}
\author{Pisin Chen}
\email{pisinchen@phys.ntu.edu.tw}
\affiliation{Graduate Institute of Astrophysics, National Taiwan University, Taipei 10617, Taiwan}
\affiliation{Department of Physics and Center for Theoretical Sciences, National Taiwan University, Taipei 10617, Taiwan}
\affiliation{Leung Center for Cosmology and Particle Astrophysics, National Taiwan University, Taipei 10617, Taiwan}
\date{\today}

\begin{abstract}
Foreground masking and incomplete sky coverage complicate CMB polarization analyses by inducing mode coupling and imperfect E/B separation, with particularly strong impact on searches for primordial $B$-modes. We present SkyReconNet-P, a convolutional neural network for inpainting CMB polarization maps that extends the SkyReconNet framework to jointly reconstruct the polarization $(Q,U)$ maps from partial-sky observations. The method combines regional processing with a hybrid design, utilizing standard convolution and dilated convolution to do a multiscale feature integration. We evaluate performance at both the map and power spectrum level using two masking scenarios: a generated random mask and the Planck 2018 common polarization inpainting mask. For both masking scenarios, SkyReconNet-P reproduces the large-scale morphology of the target maps. In power-spectrum space, we find that the reconstructed $E$-mode spectrum closely tracks the target at low multipoles, while small biases emerge at higher $\ell$. For $B$-mode, the raw reconstructed spectra exhibit a larger multipole-dependent bias, which we mitigate using a simulation-based linear calibration. We show that the calibrated $B$-mode spectrum preserve more information by comparing it with spectrum estimation using pseudo-$C_\ell$. Finally, we demonstrate cosmological parameter inference from calibrated reconstructed spectra by fitting $(r, A_{\rm lens})$ with a Gaussian bandpower likelihood, recovering posteriors consistent with injected parameters across three test ensembles down to $r \sim 10^{-3}$. These results support inpainting as a complementary route to cut-sky approaches when downstream pipelines can greatly benefit from statistically well-characterized, gap-filled polarization maps.
\end{abstract}

\maketitle

\section{Introduction}

Precision measurements of the Cosmic Microwave Background (CMB) radiation have emerged as a cornerstone of modern cosmology, providing critical constraints on the physics of the early Universe. Over the years, various satellite~\cite{Planck:2018nkj,bennett2013nine}, ground-based~\cite{thornton2016atacama,BICEP:2021xfz,BICEP2:2014owc},  and balloon-borne~\cite{ANITA:2019wyx} CMB experiments have delivered high-sensitivity CMB temperature and polarization data. These observations played a significant role in advancing our understanding of early-universe physics through both temperature and polarization anisotropies. There are several planned next-generation efforts including AliCPT~\cite{Li:2017drr}, the Simons Observatory~\cite{SimonsObservatory:2018koc}, CMB-S4~\cite{Abazajian:2019eic}, and LiteBIRD~\cite{LiteBIRD:2022cnt} are planned to observe CMB at even higher sensitivity, broader frequency coverage, and lower instrumental noise, promising unprecedented 
insights into the CMB. These data sharpen tests of early-universe physics through both temperature anisotropies and, crucially, polarization.

A particularly compelling target for CMB analysis is the detection of signatures of primordial $B$-mode polarization. Unlike lensing-induced $B$-modes produced due to gravitational lensing of the CMB photons due to large-scale structure, primordial $B$-modes are sourced by tensor metric perturbations through Thomson scattering at recombination, as such a detection would provide an indirect probe of Primordial Gravitational Waves (PGWs) produced during inflation~\cite{Seljak:1996gy,Kamionkowski:1997av}. Primordial $B$-modes also offer a way to discriminate between competing inflationary models~\cite{CMBPolStudyTeam:2008rgp,Planck:2018jri}, since the tensor-to-scalar ratio $r$ (inferred from a primordial $B$-mode measurement) could constraint to the tensor amplitude and, in the slow-roll inflation, the slow-roll parameters, thereby constraining the energy scale and shape of the inflationary potential.

Despite major experimental advances, primordial B modes remain undetected; recent analyses place an upper limit of $r < 0.032$ (95\% C.L.)~\cite{Tristram:2021tvh}. Achieving a robust measurement is challenging because the primordial signal is extremely faint and must be separated from several contaminants, including Galactic foreground emission, instrumental noise, and lensing-induced B modes, as well as analysis-induced mixing between E and B polarization. In particular, incomplete sky coverage breaks the ideal orthogonality of E and B modes and induces E-to-B leakage, allowing the much larger E-mode signal to contaminate $B$-mode estimates~\cite{Bunn:2002df,Smith:2006hv}. Incomplete coverage is unavoidable in practice: regions with strong synchrotron, thermal dust, and point sources are usually excluded using foreground masks, especially near the Galactic plane~\cite{Planck:2018yye}, leaving cut-sky polarization maps for cosmological analysis.

Power spectra on a masked sky are commonly estimated using fast pseudo-$C_\ell$ approaches (MASTER~\cite{Hivon:2001jp,Alonso:2018jzx,Smith:2006hv}) or near–optimal quadratic maximum-likelihood (QML) estimators~\cite{Tegmark:1996qt,Tegmark:2001zv}. While these methods are mature and widely used, masking inevitably induces mode coupling and imperfect E/B separation while working with CMB polarization data, which can complicate inference for $B$-mode searches and related analyses~\cite{Bunn:2002df,Bunn:2016lxi}.  

Beyond power-spectrum estimation, statistically consistent “filled” maps with reduced boundary artifacts can be benefit for downstream analyses. For example, low-$\ell$ (large-angular-scale) analyses and anomaly tests are particularly sensitive to sky cuts and mode coupling, and have motivated full-sky inpainting treatments~\cite{starck2013low}. Likewise, CMB lensing pipelines commonly need to mitigate biases from compact-source and cluster masking; in practice, small masked holes are often filled using Gaussian constrained realizations prior to lensing reconstruction or map filtering~\cite{bucher2012filling, Benoit-Levy:2013uxb, Darwish:2020fwf}. More broadly, analyses involving non-local statistics (e.g.\ certain non-Gaussianity or isotropy tests) are often simplest to formulate on full-sky maps, motivating efficient and well-characterized gap-filling schemes~\cite{bucher2012filling,Abrial:2008mz}.

An alternative approach that follows this path is \emph{inpainting}, rather than working directly with a cut-sky map, one reconstructs masked pixels using information from the surrounding observed region under an assumed prior or model. 
Classical approaches include Gaussian constrained realizations~\cite{hoffman1991constrained,bucher2012filling} and harmonic-/sparsity-domain inpainting on the sphere~\cite{Abrial:2008mz,starck2013low}. More recently, machine-learning-based inpainting has been explored, including U-Net–style architectures with partial convolutions~\cite{Montefalcone:2020fkb}, as well as GAN~\cite{Sadr:2020rje} and VAE~\cite{Yi:2020xgq}. 


Recently, we introduced SkyReconNet~\cite{Lambaga:2025nbw}, a multiscale feature-integration architecture combining dilated and standard convolutions for reconstructing CMB temperature maps. The network accurately recovered both large-scale structure and small-scale fluctuations in masked regions, leveraging its hybrid design. In this work, we introduce SkyReconNet-P, a natural extension of our previous framework~\cite{Lambaga:2025nbw} to the more challenging problem of CMB polarization map reconstruction. We adapt the SkyReconNet architecture for joint reconstruction of CMB polarization maps from partial-sky maps and evaluate the resulting maps in downstream analyses. In particular, we perform parameter estimation using power spectra derived from reconstructed maps, showing that inpainting provides a viable and potentially better alternative for traditional methods for certain use cases.

The remainder of this paper is organized as follows.  Section~\ref{sec:NetArch} details the network architecture and loss functions for CMB polarization map reconstruction. Then in section~\ref{sec:dataset} we outline our dataset generation procedure for CMB simulation and the mask map. Afterward we describe our methodology and training procedure in section~\ref{sec:methodology}. The reconstruction results and associated analyses are reported in section~\ref{sec:Results}. In section~\ref{sec:parameter}, we demonstrates cosmological parameter estimation using reconstructed data. We conclude in Section~\ref{sec:Conclusions} with a discussion of future directions and prospects for application to forthcoming CMB experiments.

\begin{figure*}
  \centering
  \includegraphics[width=0.9\textwidth]{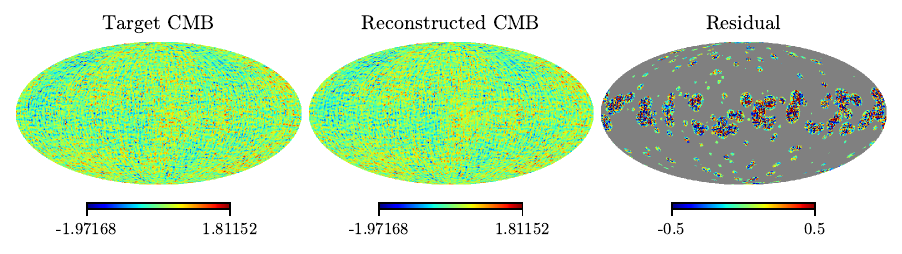}\par
  \vspace{0.5em}
  \includegraphics[width=0.9\textwidth]{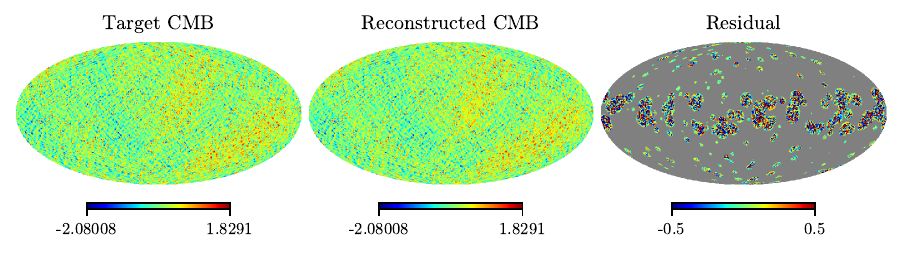}
  \caption{
  Three-panel view of $Q$ (top) and $U$ (bottom) polarization maps from the trained network with the generated mask. From left to right: target (unmasked) map, reconstructed map, and the residual (reconstructed minus target).
  }
  \label{fig:map_1}
\end{figure*}

\begin{figure*}
  \centering
  \includegraphics[width=0.9\textwidth]{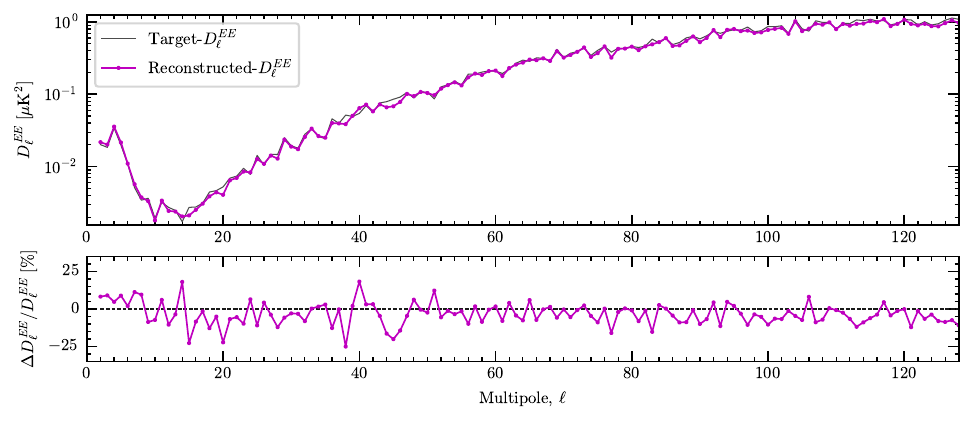}
  \caption{
  E-mode angular power spectrum for the trained network with the generated mask. The upper panel shows $D_\ell^{EE}$ from the target (grey) and reconstructed (coloured) maps on the masked sky. The lower panel shows the percentage difference between the two. The reconstruction tracks the target spectrum closely at low multipoles, with an increasing negative bias for $\ell \gtrsim 60$.
  }
  \label{fig:Dl_EE_1}
\end{figure*}

\begin{figure*}
  \centering
  \includegraphics[width=0.9\textwidth]{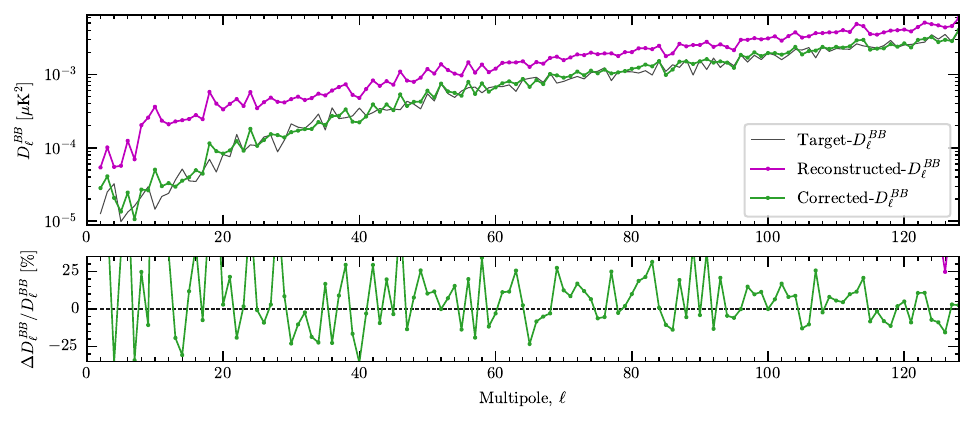}
  \caption{
  $B$-mode angular power spectrum for the same realization ($r = 0.00022$). The upper panel compares the target $D_\ell^{BB}$, the raw reconstructed spectrum (magenta), and the BB-calibrated spectrum (green) obtained using the amplitude-correction model described in the text. The lower panel shows the corresponding percentage differences with respect to the target. The calibration substantially reduces the mean bias in $D_\ell^{BB}$ over the analysis range, leaving predominantly realization-to-realization fluctuations.
  }
  \label{fig:Dl_BB_1}
\end{figure*}

\section{Network Architecture}\label{sec:NetArch}

SkyReconNet-P builds on SkyReconNet~\cite{Lambaga:2025nbw}, a hybrid Convolutional Neural Network (CNN) that combines standard and dilated convolutional operations to inpaint missing pixels in CMB temperature maps. The overall workflow—including regional partitioning, nested U-Net processing, and stitching—follows the approach implemented in~\cite{Lambaga:2025nbw}, with two practical modifications for polarization: (i) joint reconstruction of $(Q,U)$ maps and (ii) replacing the Mean Squared Error (MSE) loss with the Mean Absolute Error (MAE) loss in the loss function, alongside the previously used Structural Similarity Index Measure (SSIM) loss. We summarize the components most relevant to this work; full architectural details are provided in~\cite{Lambaga:2025nbw}.

\subsection{Network}
The SkyReconNet-P architecture follows the division strategy of~\cite{Sudevan:2024hwq,Lambaga:2025nbw}, we split each full-sky map into $n=4$ planar regions of size $3N_{side}\times N_{side}$ using the HEALPix~\cite{Gorski:2004by} pixelization scheme. Each region is processed by a dedicated U-Net sub-network~\cite{ronneberger2015u}, and the outputs are recombined into a full-sky prediction.

Each sub-network is a four-level encoder-decoder with skip connections. Convolution blocks combine standard and dilated convolutions with residual connections to enlarge the receptive field while retaining fine-scale structure. We use Batch Normalization and \textit{p}-ReLU activations~\cite{he2015delving}. In the decoder, feature maps are upsampled and fused with the corresponding encoder features via skip connections.

To process polarization, the network input stacks the masked $Q$ and $U$ maps together with the binary mask (three channels), and the network predicts $(Q,U)$ jointly (two channels). We apply a fixed normalization before the regional split and invert it after recombination. Finally, we replace only masked pixels with the network prediction, leaving the observed region unchanged.

\subsection{Loss Function}\label{sec:loss}
We employ a composite loss similar to~\cite{Lambaga:2025nbw}, combining a MAE and a SSIM term~\cite{SSIM2004}. The MAE term is
\begin{equation}
    \mathcal{L}_{\mathrm{MAE}} = \frac{1}{N_i N_j} \sum_{i,j}^{N_i,N_j} \left| \mathbf{Y}_{ij} - \hat{\mathbf{Y}}_{ij} \right|,
\end{equation}
where $\mathbf{Y}$ and $\hat{\mathbf{Y}}$ denote the ground-truth and predicted images, and $N_i$ and $N_j$ are the image width and height.

The SSIM-derived loss is
\begin{equation}
    \mathcal{L}_{\mathrm{SSIM}} = 1 - \frac{\left(2\mu_{\mathbf{Y}}\mu_{\hat{\mathbf{Y}}} + c_1\right)\left(2\sigma_{\mathbf{Y}\hat{\mathbf{Y}}} + c_2\right)}{\left(\mu_{\mathbf{Y}}^2 + \mu_{\hat{\mathbf{Y}}}^2 + c_1\right)\left(\sigma_{\mathbf{Y}}^2 + \sigma_{\hat{\mathbf{Y}}}^2 + c_2\right)},
\end{equation}
with $\mu$ and $\sigma$ denoting the (local) mean and variance of $\mathbf{Y}$ or $\hat{\mathbf{Y}}$, and $\sigma_{\mathbf{Y}\hat{\mathbf{Y}}}$ their covariance. We set $c_1$ and $c_2$ as 0.01 and 0.03, respectively. 

The overall objective is
\begin{equation}
    \mathcal{L} = \alpha\, \mathcal{L}_{\mathrm{MAE}} + \beta\, \mathcal{L}_{\mathrm{SSIM}}\,,
\end{equation}
where $\alpha$ and $\beta$ control the relative influence of pixel-level accuracy and structural similarity.

\begin{figure*}
  \centering
  \includegraphics[width=0.9\textwidth]{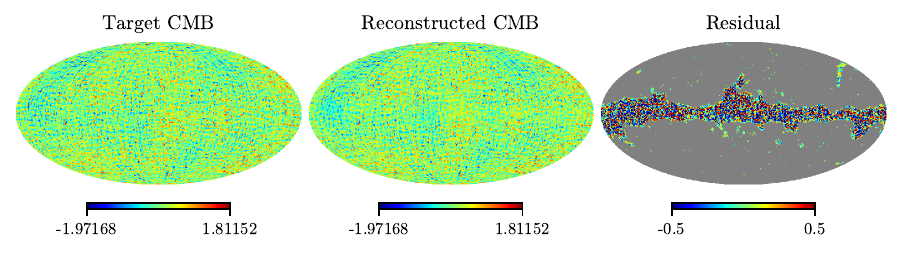} \par
  \vspace{0.5em}
  \includegraphics[width=0.9\textwidth]{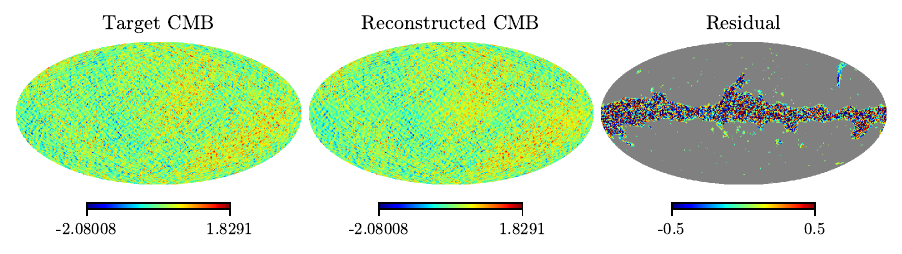}
  \caption{
  Three-panel view of $Q$ (top) and $U$ (bottom) polarization maps from the trained network with the Planck mask. From left to right: target (unmasked) map, reconstructed map, and the residual (reconstructed minus target).
  }
  \label{fig:map_2}
\end{figure*}

\begin{figure*}
  \centering
  \includegraphics[width=0.9\textwidth]{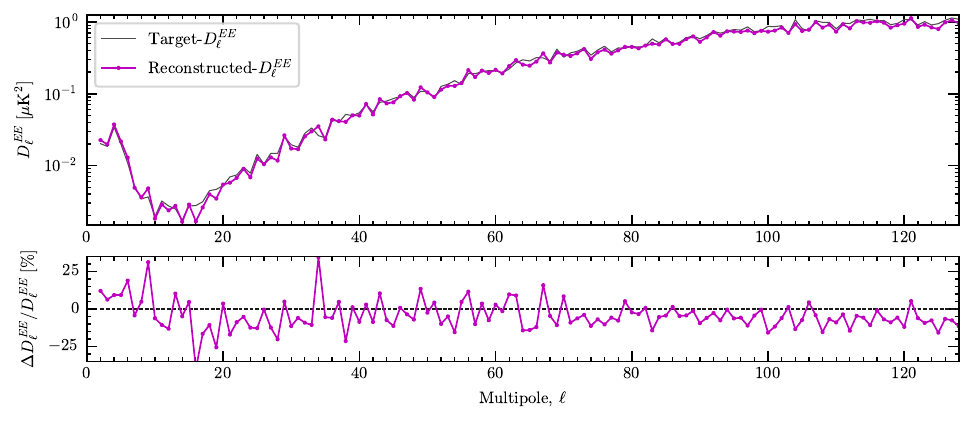}
  \caption{
  E-mode angular power spectrum for the trained network with the Planck mask. The upper panel shows $D_\ell^{EE}$ from the target (grey) and reconstructed (coloured) maps on the masked sky, and the lower panel shows the percentage difference between the two. The reconstruction tracks the target spectrum closely at low multipoles, with an increasing negative bias for $\ell \gtrsim 60$.
  }
  \label{fig:Dl_EE_2}
\end{figure*}

\begin{figure*}
  \centering
  \includegraphics[width=0.9\textwidth]{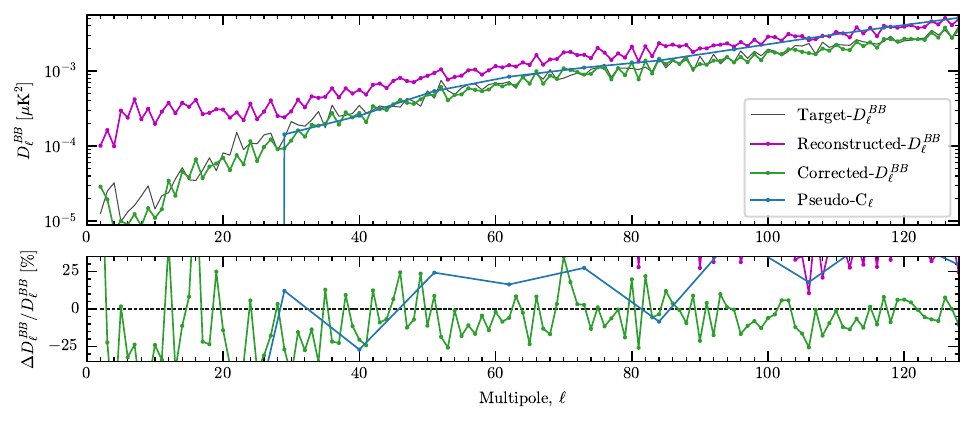}
  \caption{
  $B$-mode angular power spectrum for the same realization ($r = 0.00022$). The upper panel compares the target $D_\ell^{BB}$, the raw reconstructed spectrum (magenta), and the BB-calibrated spectrum (green) obtained using the amplitude-correction model. The lower panel shows the corresponding percentage differences with respect to the target. The calibration reduces the mean bias in $D_\ell^{BB}$ and improves agreement with the target across the plotted multipole range. The comparison with Pseudo-$C_\ell$ is also shown in blue, using apodization and binning with bin size 11.
  }
  \label{fig:Dl_BB_2}
\end{figure*}

\section{Dataset Generation}\label{sec:dataset}

We generate simulated CMB polarization maps and construct masked inputs following the preprocessing pipeline of~\cite{Lambaga:2025nbw,Sudevan:2024hwq}, extended to polarization. We summarize the steps required to reproduce our training and evaluation datasets.

\subsection{CMB Simulations}
We adopt the $\Lambda$CDM model, fixing all cosmological parameters to their Planck 2018 best-fit values~\cite{Planck:2018vyg} except for $r$ and $A_{\rm lens}$. Specifically, we use
$H_0 = 67.37~km~s^{-1}~Mpc^{-1}$,
$\Omega_b h^2 = 0.02233$, $\Omega_c h^2 = 0.1198$, $\tau = 0.0540$,
$10^9 A_s = 2.105$ and $n_s = 0.9652$,
and sample $A_{\rm lens}$ uniformly on $[0.9,1.1]$ and $r$ from a log-uniform distribution from $[1 \times 10^{-4}, 5.5 \times 10^{-2}]$.
For each parameter set, CAMB~\cite{lewis2000efficient,lewis2011camb} is used to compute angular power spectra, from which we generate HEALPix polarization $Q$ and $U$ maps at $\mathrm{NSIDE} = 64$ and smooth them with a Gaussian beam of 0.92$^\circ$ FWHM.

To interface with the CNN, we convert each HEALPix map from RING to NESTED ordering and reshape each full-sky map into four planar images of size $3N_{side}\times N_{side}$ as in~\cite{Lambaga:2025nbw,Sudevan:2024hwq}. Inputs are constructed by applying the corresponding planar mask to the $Q$ and $U$ images and appending the planar mask as an additional channel. Targets are the unmasked planar $Q$ and $U$ maps. 
In total we used 3,072 sets of maps, where we reserve 512 sets for testing and 512 for validation; the remaining 2,048 sets are used for training.

\subsection{Masks}\label{subsec:masks}
We train separate models for two masking scenarios. The first is a synthetic random mask consisting of 250 circular holes: 150 holes are placed within $\pm 30^{\circ}$ latitude with radii up to 10 pixels, and 100 holes are placed at higher latitudes with radii up to 4 pixels, yielding a masked sky fraction of 21\%. The second mask is the Planck 2018 common polarization inpainting mask~\cite{Planck:2018yye}, which excludes pixels with low component-separation confidence, the Galactic plane, and point sources.

\section{Methodology}\label{sec:methodology}
For each mask, we train SkyReconNet-P using the AdamW optimization scheme \cite{loshchilov2017decoupled} in a two-stage training procedure. In the first stage, we train the network for 100 epochs with batch size 64, initial learning rate 0.012, and weight decay 0.09. We use a cosine learning-rate schedule with 5-epoch warmup and a minimum learning rate of $3\times 10^{-6}$. In the second stage, we resume training from the best checkpoint from the first stage and continue for an additional 1000 epochs. The loss in Sec.~\ref{sec:loss} is used with weights $\alpha = 0.5$ and $\beta = 1.15$. the MAE loss is applied to both $Q$ and $U$ maps, while the SSIM loss is applied to $P^2 = Q^2 +U^2$, with SSIM window size set to 5. We evaluate performance on the test set by comparing reconstructed and ground-truth polarization maps and their $E$-mode and $B$-mode power spectra.

\section{Results}\label{sec:Results}

We evaluate SkyReconNet-P on the testing dataset for two masking scenarios: (i) a synthetic “generated” mask with randomly distributed holes, and (ii) the Planck 2018 common polarization inpainting mask~\cite{Planck:2018yye}, described in the section~\ref{subsec:masks}. We summarize both the map-level reconstructions and the polarization power spectra.

\subsection{Generated Mask}\label{subsec:ResGenMask}

Figure~\ref{fig:map_1} shows a sample of $Q$ and $U$ maps reconstruction using the trained network with the generated mask map. In both cases, the reconstructed maps reproduce the large-scale morphology of the target while avoiding obvious boundary artifacts around the masked regions. The residual maps are dominated by localized structure within the masked holes rather than coherent large-scale patterns.

We quantify this agreement via the polarization power spectra in Figs.~\ref{fig:Dl_EE_1} and~\ref{fig:Dl_BB_1}. For $E$-modes, we see a very close agreement between the reconstructed and the target spectrum, as evidenced by the percentage difference plot in the second panel of Fig.~\ref{fig:Dl_EE_1}. For $B$-modes, the raw reconstructed spectrum exhibits a stronger bias, motivating a calibration step. We therefore apply the BB calibration procedure of~\cite{Sudevan:2025ncj} using a reserved calibration set with $N_{\rm cal}=60$ realizations. For each realization $s$ and multipole $\ell$, we define
\begin{eqnarray}
    R_s(\ell)&=& \frac{D_{\ell,s}^{BB,\rm true}}{D_{\ell,s}^{BB,\rm pred} + \varepsilon}, \\
    A_s &=& \left\langle D_{\ell,s}^{BB,\rm pred} \right\rangle_{30 \le \ell \le 120},
\end{eqnarray}
where $\varepsilon$ is a small regularization constant. For each $\ell$ we fit a linear relation
\begin{equation}
R_s(\ell) \simeq m(\ell) A_s + c(\ell),
\end{equation}
and apply it to a predicted spectrum via $D_{\ell,s}^{BB,\rm corr} = [m(\ell) A_s' + c(\ell)]\,D_{\ell,s}^{BB,\rm pred}$, where $A_s'$ is computed from the predicted spectrum in the same multipole window. As shown in Fig.~\ref{fig:Dl_BB_1}, this correction brings the mean reconstructed $B$-mode spectrum into substantially closer agreement with the target over the calibration range while leaving the stochastic scatter largely unchanged.

\subsection{Planck Mask}

We repeat the same evaluation for the trained network with the Planck 2018 common polarization inpainting mask, which introduces a large, structured sky cut. Figure~\ref{fig:map_2} shows a reconstruction of a sample from the testing dataset. The residuals are concentrated in the masked region, as expected for a Galactic-plane-dominated cut, while the reconstructed maps preserve the qualitative structure of the target over the available sky.

The corresponding spectra are shown in Figs.~\ref{fig:Dl_EE_2} and~\ref{fig:Dl_BB_2}. The reconstructed $E$-mode spectrum follows the target closely, with a small but noticeable negative bias at higher multipoles. For $B$-modes, we apply the same calibration strategy as outlined in the subsection~\ref{subsec:ResGenMask}; after calibration, the corrected spectrum tracks the target significantly better across the analysis range, with remaining discrepancies dominated by multipole-dependent fluctuations. 
We also compare against a pseudo-$C_\ell$ estimate computed with NaMaster~\cite{Alonso:2018jzx}. We used $3^\circ$ $C2$ apodization and bandpower binning with $\Delta \ell = 11$ while computing full-sky $C-\ell$ using NaMaster algorithm. In the lower multipole bins ($\ell \lesssim 30$), the pseudo-$C_\ell$ estimate can become negative.
By contrast, the calibrated $B$-mode spectrum tracks the target spectrum more closely across the plotted range, even without additional binning, indicating improved preservation of the overall $B$-mode spectral shape and amplitude for this analysis setup.

\begin{figure}
  \centering
  \includegraphics[width=0.48\textwidth]{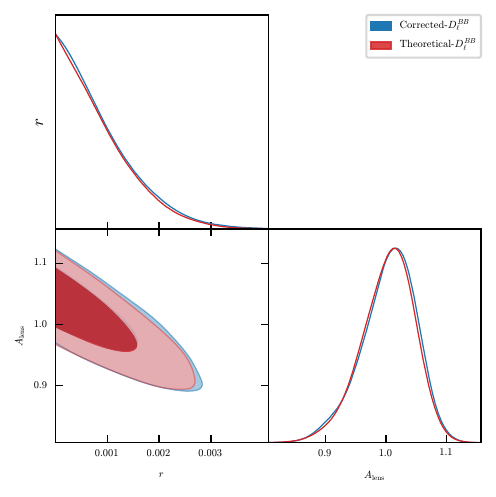}
  \caption{Null-test posterior distributions for the tensor-to-scalar ratio $r$ and the lensing amplitude $A_{\rm lens}$ using a fixed simulated dataset with $r=0$ and $A_{\rm lens}=1.0$. The corrected spectrum null test is indicated with blue color, while the theoretical null test is shown in red. This result demonstrates that our network does not spuriously recover a primordial $B$-mode signal in the null case.}
  \label{fig:corner_r_Alens_null}
\end{figure}

\begin{figure}
  \centering
  \includegraphics[width=0.48\textwidth]{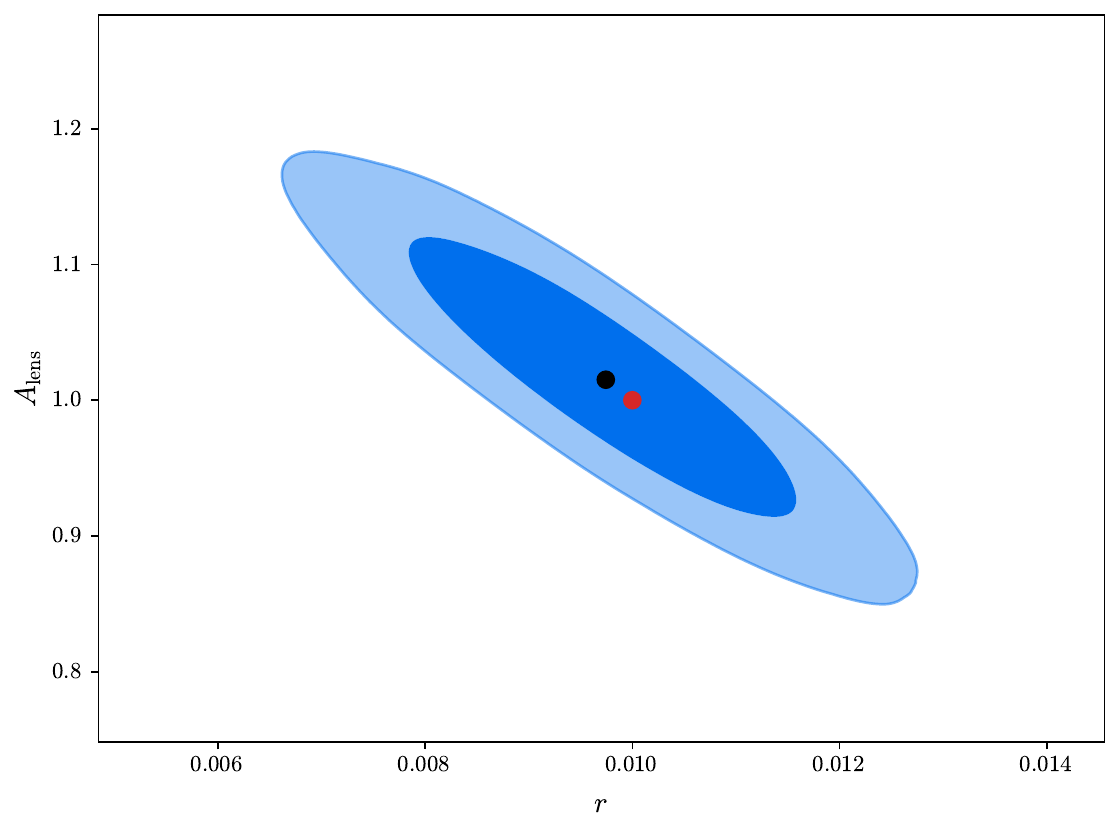}
  \includegraphics[width=0.48\textwidth]{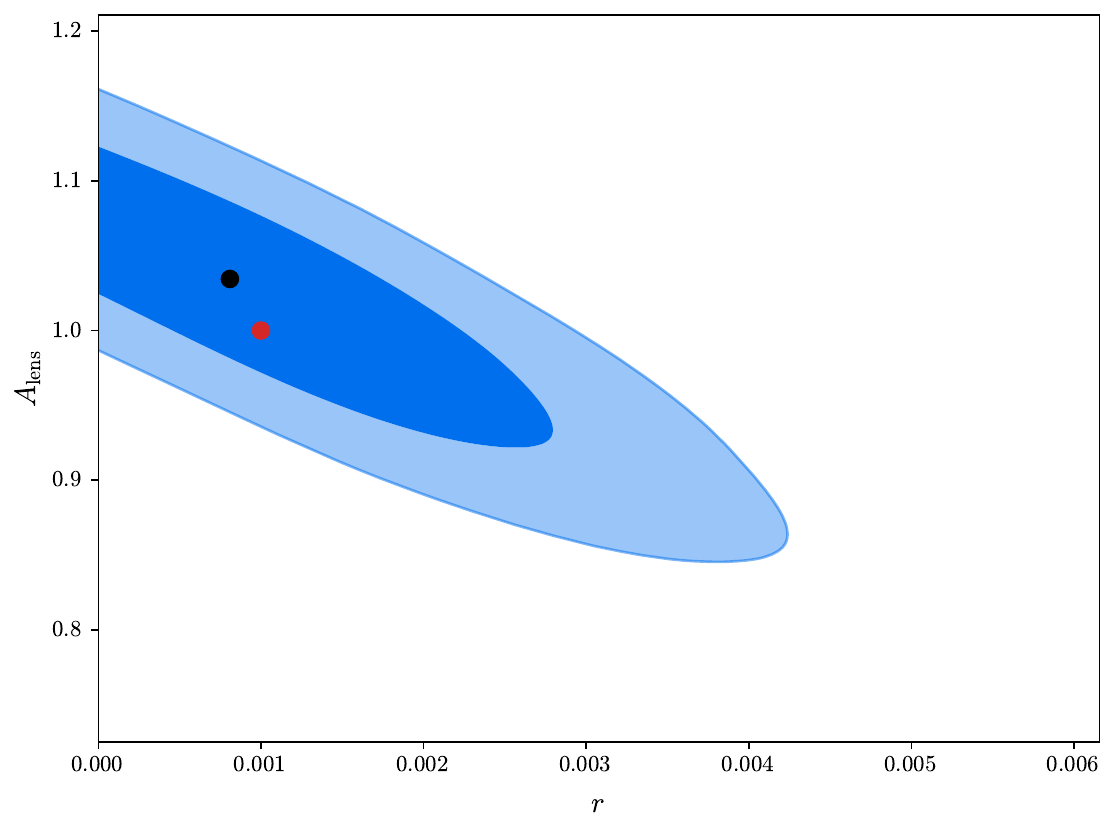} 
  \caption{Joint posterior distributions for the tensor-to-scalar ratio $r$ and the lensing amplitude $A_{\rm lens}$ for the two ensemble datasets with input $r=0.01$ (top) and $r=0.001$ (bottom), all with $A_{\rm lens}=1$, using the Planck mask map model. Contours show the 68\% and 95\% credible regions, the black dots mark the peak of posterior distributions while the red dots mark the true parameter values.}
  \label{fig:corner_r_Alens}
\end{figure}

\begin{figure*}
  \centering
  \includegraphics[width=0.98\textwidth]{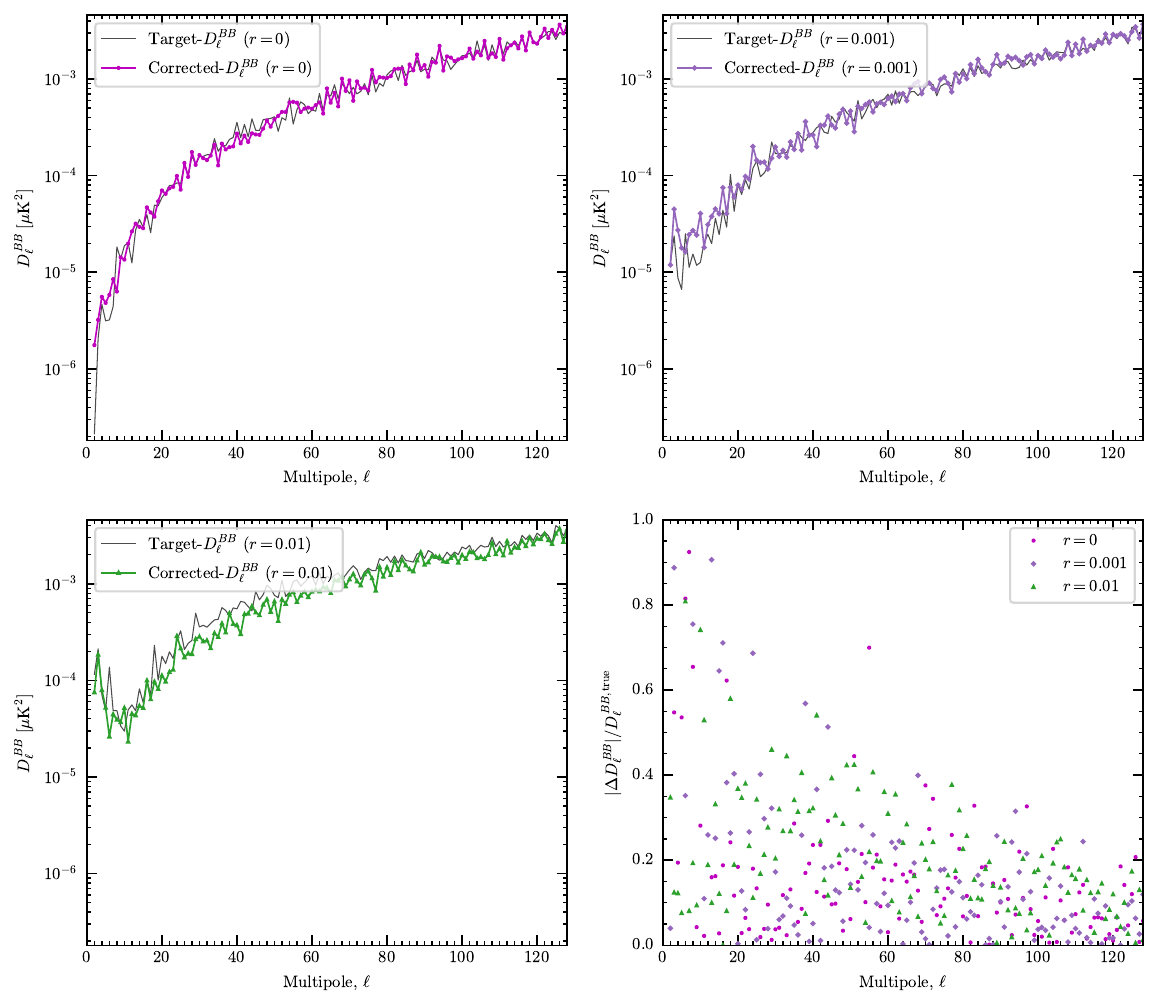}
  \caption{$B$-mode spectrum summary. Top and bottom left panels: target vs corrected $B$-mode spectrum from one realization of the null case ensemble ($r=0$, top left), and  non-null case ensembles ($r = 0.001$ on the top right and $r = 0.01$ on the bottom left). Bottom right panel: fractional absolute residual of all spectra from each ensembles.}
  \label{fig:CPE_BB}
\end{figure*}

\begin{figure}
  \centering
  \includegraphics[width=\columnwidth]{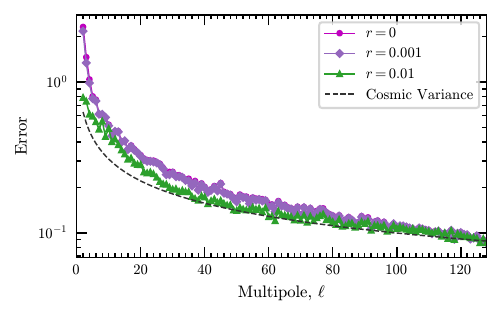}
  \caption{Error of corrected $B$-mode spectra from each ensembles, by taking the standard deviation from 500 samples, compared with the cosmic-variance (dashed line).}
  \label{fig:CPE_BB_std}
\end{figure}

\begin{figure}
  \centering
  \includegraphics[width=\columnwidth]{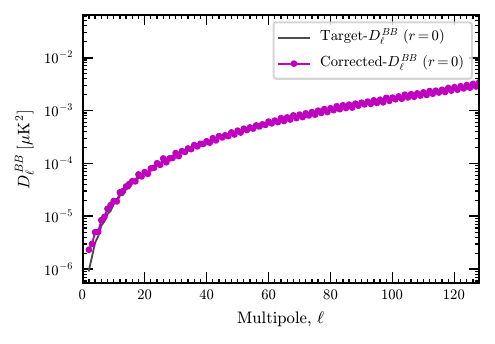}
  \caption{Comparison of target and corrected mean $B$-mode spectra from 500 samples in the null case ($r=0$) ensemble. Gray curves show target spectra while colored marker-line show corrected predictions.}
  \label{fig:CPE_BB_mean}
\end{figure}

\section{Parameter estimation}\label{sec:parameter}

We assess how well the reconstructed polarization maps can be used for cosmological parameter inference. In this section we focus on the tensor-to-scalar ratio $r$ and the lensing amplitude $A_{\rm lens}$, which together control the amplitude of $B$-mode polarization.

For parameter estimation we first generate dedicated ensembles of simulations at fixed cosmology. Each ensemble consists of masked Q/U maps at $\mathrm{NSIDE}=64$ with a single choice of the tensor amplitude. The first ensemble is a null test dataset, with a fixed cosmology of $r = 0$ and $A_{\rm lens} = 1.0$. The other ensembles have $r = 0.01$ and $0.001$, while keeping $A_{\rm lens}=1$ fixed. The other cosmological parameters are fixed to the Planck 2018 best-fit values~\cite{Planck:2018vyg}. The maps are processed with the trained neural network from the previous section, using the Planck mask variant. For every realization we compute the $B$-mode power spectrum from the reconstructed maps using the same mask, beam, and pixel-window treatment as in the previous section. We then apply the BB linear calibration, so that the data vector entering the likelihood is a set of corrected bandpowers $\hat{X}_b$ constructed from the corrected $D_\ell^{BB}$. 

Fig.~\ref{fig:CPE_BB} summarizes representative corrected spectra and residuals for the three ensembles. Across ensembles, the corrected spectra closely track their respective target spectra and clearly separate according to the injected $r$, most notably at low multipoles. The fractional absolute residuals of each spectra also remain small across most multipoles and are generally largest at low $\ell$, where cosmic variance and noise have the strongest impact.

Additionally in Fig.~\ref{fig:CPE_BB_std} we plot the uncertainty of the corrected $B$-mode spectra, estimated as the standard deviation over 500 realizations for each ensemble, and compare it to the cosmic-variance limit. The error decreases with increasing multipole and is broadly consistent with being near the cosmic-variance floor at high $\ell$, while being larger at low multipole $\ell < 30$ where fluctuations are intrinsically larger. Complementarily, Fig.~\ref{fig:CPE_BB_mean} shows that in the null case ($r=0$) the corrected mean spectrum closely matches the target mean across the full multipole range, indicating that the correction is effectively unbiased at the level of the ensemble average.

The spectra are compressed into $N_{\rm bin}=18$ bandpowers between $\ell=30$ and $\ell=120$ with approximately uniform width $\Delta\ell \simeq 5$. The binning is defined by simple top-hat windows in multipole space; all three ensembles are reduced to the same bandpower definition, so that they differ only by the underlying input value of $r$.

We adopt a Gaussian likelihood,
\begin{equation}
  -2 \ln \mathcal{L} =
  (\hat{\mathbf X} - \mathbf X^{\rm th})^{\mathsf T}
  M_{\rm fid}^{-1}
  (\hat{\mathbf X} - \mathbf X^{\rm th}),
\end{equation}
where $\hat{\mathbf X}$ denotes the vector of measured BB bandpowers $\{\hat{X}_b\}$ from the reconstructed maps, $\mathbf X^{\rm th} \equiv \mathbf X^{\rm th}(r, A_{\rm lens})$ is the corresponding theoretical bandpower vector for a given $(r, A_{\rm lens})$, and $M_{\rm fid}$ is a fiducial bandpower covariance matrix. In this formulation the likelihood penalizes departures of the data from the theoretical model in units of the expected covariance; $M_{\rm fid}$ is kept fixed while varying $(r, A_{\rm lens})$, which is an excellent approximation in the low-signal regime considered here.

The theoretical BB spectrum is modeled analytically instead of recomputing CAMB for every likelihood evaluation. We precompute two CAMB templates at $\mathrm{NSIDE}=64$: a tensor template with $r = r_{\rm pivot} = 0.03$ and no lensing ($A_{\rm lens}=0$), and a purely lensing template with $r=0$ and $A_{\rm lens}=1$. For arbitrary $(r, A_{\rm lens})$ we then form
\begin{equation}
  C_\ell^{BB}(r, A_{\rm lens}) =
  \frac{r}{r_{\rm pivot}}\,C_{\ell}^{\rm tensor}
  + A_{\rm lens}\,C_{\ell}^{\rm lensing},    
\end{equation}
and convert to $D_\ell^{BB}$ and to bandpowers using the same binning as for the data. This defines the analytic model for the BB spectrum used in the inference.

To account for residual reconstruction and filtering effects we apply an $\ell$-dependent transfer function. The transfer function $T_\ell^{BB}$ is estimated by generating a large number of mock unmasked CMB signal with $r$ drawn from a log-uniform distribution in the range $[5\times10^{-4}, 5\times 10^{-2}]$ and fixed $A_{\rm lens}=1$. For each realization we compute the deconvolved $D_\ell^{BB}$ from the maps, using the same pipeline we use in the dataset construction. We then form the ratio to the corresponding analytic CAMB spectrum and the final $T_\ell^{BB}$ is the mean of these ratios at each multipole. In the likelihood we multiply the theoretical bandpowers by the binned version of this transfer function so that $X_b^{\rm th}$ includes the same response as the reconstructed spectra.

The covariance matrix $M_{\rm fid}$ is obtained from an independent fiducial simulation set with $r_{\rm fid}=0.03$ and $A_{\rm lens}=1$. Using 1000 realizations of the corrected $D_\ell^{BB}$ spectra from this fiducial run, we estimate the sample covariance of the binned bandpowers in the range $30 \le \ell \le 120$. In the main analysis we retain only the diagonal elements of this covariance matrix, i.e. we treat the bandpowers as statistically independent with variances taken from the fiducial ensemble.

We explore the posterior with Cobaya's Metropolis–Hastings sampler, varying only $r$ and $A_{\rm lens}$ and keeping all other cosmological parameters fixed to a Planck $\Lambda$CDM model. The priors on $r$ and $A_{\rm lens}$ are a uniform distributions with allowed range $0 \le r \le 0.2$ and $0 \le A_{\rm lens} \le 3$. 

With the null test ensemble we find that no evidence of primordial $B$-mode signal and obtain marginalized posterior mean for $r = 0.00085 \pm 0.00069 $ and a 95\% upper limit of $r < 2.22\times 10^{-3}$, with the lensing amplitude $A_{\rm lens} = 1.004 \pm 0.045$, consistent with the input value. As a consistency check, we repeat the analysis using the target spectra generated from the unmasked maps, rather than the corrected reconstructed spectra. From this setup we obtain mean for $r = 0.00083 \pm 0.00067 $ and the lensing amplitude $A_{\rm lens} = 1.003 \pm 0.045$. The resulting $r$ and $A_{\rm lens}$ posteriors for both setups can be seen in Fig.~\ref{fig:corner_r_Alens_null}, where the blue and red colors indicate posterior distributions from corrected and theoretical null tests, respectively. This agreement indicates that, in the absence of any primordial $B$-mode signal, our reconstruction and inference procedure does not spuriously recover one. 

Using the next ensemble with $r=0.01$ and $A_{\rm lens} = 1.0$, we recover
\begin{equation}
  r = 0.00971 \pm 0.00124, \qquad
  A_{\rm lens} = 1.017 \pm 0.068,
\end{equation}

which is a very close agreement with the true parameters. Lastly in the next ensemble case we obtain
\begin{equation}
  r = 0.00150 \pm 0.00102, \qquad
  A_{\rm lens} = 0.9994 \pm 0.0632,
\end{equation}

with the true parameters $r=0.001$ and $A_{\rm lens} = 1.0$.
In all cases the peak of the posterior is consistent with the input values and lies well within the 68\% credible interval. The corresponding two-dimensional posterior contours are shown in Fig.~\ref{fig:corner_r_Alens}. These results indicate that, after calibration, the reconstructed polarization maps preserve enough information to provide unbiased constraints on $r$ and $A_{\rm lens}$ down to $r \simeq 10^{-3}$ for the noise level and sky coverage considered here.

\section{Conclusions}\label{sec:Conclusions}

We presented SkyReconNet-P, a neural-network inpainting framework for CMB polarization that extends SkyReconNet~\cite{Lambaga:2025nbw} to jointly reconstruct the $Q$ and $U$ polarization fields on a masked sky. The method follows a regional, multiscale encoder-decoder design and incorporates a normalization strategy to stabilize training, producing full-sky completions that replace only the masked pixels while leaving the observed sky unchanged.

Using simulated polarization maps at $\mathrm{NSIDE}=64$, we evaluated SkyReconNet-P for two masking configurations: a generated random mask and the Planck 2018 common polarization inpainting mask. At the map level, reconstructions reproduce the qualitative large-scale morphology of the target maps and exhibit residuals concentrated within the masked regions. In harmonic space, the reconstructed $E$-mode spectrum matches the target well at most of the multipoles range, while a small negative bias becomes visible at higher $\ell$. For $B$-mode spectrum, the raw reconstructed spectra show a larger, multipole-dependent bias. we mitigate this effect with a simulation-based linear calibration that substantially reduces the mean discrepancy with respect to the target spectrum over the analysis range. For comparison we also compare our result with a $B$-mode power spectrum estimation using Pseudo-$C_\ell$ and we observed our calibrated spectrum fits the target spectrum more closely, preserving more information in the overall $B$-mode spectrum.

We further tested the utility of the reconstructed maps for cosmological inference by performing likelihood-based parameter estimation for the tensor-to-scalar ratio $r$ and lensing amplitude $A_{\rm lens}$ using calibrated $B$-mode bandpowers for the Planck-mask configuration. Across three ensembles with fixed $(r, A_{\rm lens})$, the recovered posteriors are consistent with the input values and support unbiased inference at the level of the quoted uncertainties down to $r \sim 10^{-3}$ for the sky coverage and pipeline assumptions adopted here.

Overall this study shows that the proposed method can be extended to CMB polarization in a way that preserves the information needed for downstream analyses. We show that inpainting can be useful as a practical complement to standard cut-sky pipelines when full-sky, gap-filled polarization maps are advantageous. More broadly, this work provides a concrete example of how machine-learning reconstructions can be made quantitatively useful for cosmological inference by validating their impact at the spectrum and parameter level, rather than only at the map level.





\begin{acknowledgments}
RDL and PC was supported by the Leung Center for Cosmology and Particle Astrophysics (LeCosPA), National Taiwan University (NTU). 
This work is based on observations obtained with Planck
(http://www.esa.int/Planck). Planck is an ESA science mission with instruments 
and contributions directly funded by
ESA Member States, NASA, and Canada. We acknowledge the use of Planck Legacy Archive (PLA). 
We use the publicly available HEALPix~\citep{Gorski:2004by} package 
(http://healpix.sourceforge.net) for the
analysis of this work. The network we have developed is
based on the libraries provided by Tensorflow, although the same
can be designed using other ML-based platforms.
\end{acknowledgments}

\bibliography{main}

\end{document}